\pdfoutput=1 

\documentclass[conference]{IEEEtran}
\usepackage{cite}
\usepackage{threeparttable}
\usepackage{amsmath,amssymb,amsfonts}
\usepackage{enumitem}
\ifCLASSOPTIONcompsoc
\usepackage[caption=false,font=normalsize,labelfon
t=sf,textfont=sf]{subfig}
\else
\usepackage[caption=false,font=footnotesize]{subfig}
\fi
\usepackage{multirow}
\usepackage[super]{nth}
\usepackage{layouts}

\usepackage{algorithmic}
\usepackage{graphicx}
\usepackage{textcomp}
\usepackage{xcolor}
\usepackage{pifont}
\usepackage{makecell}
\def\BibTeX{{\rm B\kern-.05em{\sc i\kern-.025em b}\kern-.08em
    T\kern-.1667em\lower.7ex\hbox{E}\kern-.125emX}}

\definecolor{green}{RGB}{0,128,0}
\definecolor{red}{RGB}{255,0,0}
\newcommand{\greentick}{\textcolor{green}{\ding{51}}}
\newcommand{\redx}{\textcolor{red}{\ding{55}}}

\usepackage{flushend}

\makeatletter
\newcommand*\titleheader[1]{\gdef\@titleheader{#1}}
\AtBeginDocument{
\let\st@red@title\@title
\def\@title{     
\bgroup\normalfont\normalsize\centering\@titleheader\par\egroup
\vskip1ex\st@red@title}
}
\makeatother

\begin{document}
\bstctlcite{IEEEexample:BSTcontrol} 
\setlength{\abovedisplayskip}{2ex}
\setlength{\belowdisplayskip}{2ex}

\title{{\vspace{-1.5cm}\small To appear at the 42nd IEEE/ACM International Conference on Computer Aided Design (ICCAD'23), San Francisco, CA, USA, 2023.\\
https://doi.org/10.1109/ICCAD57390.2023.10323613\\ }\vspace{-0.8\baselineskip}
\rule{\textwidth}{0.4pt}
Bespoke Approximation of Multiplication-Accumulation and Activation Targeting Printed Multilayer Perceptrons \vspace{-1.1ex}
}


\author{\IEEEauthorblockN{
Florentia Afentaki\IEEEauthorrefmark{1}\IEEEauthorrefmark{3},
Gurol Saglam\IEEEauthorrefmark{3},
Argyris Kokkinis\IEEEauthorrefmark{4},
Kostas Siozios\IEEEauthorrefmark{4},
Georgios Zervakis\IEEEauthorrefmark{1},
Mehdi B. Tahoori\IEEEauthorrefmark{3}
}
\IEEEauthorblockA{
\IEEEauthorrefmark{1}University of Patras, Greece,
\IEEEauthorrefmark{4}Aristotle University of Thessaloniki, Greece,
\IEEEauthorrefmark{3}Karlsruhe Institute of Technology, Germany
}
\IEEEauthorblockA{
\IEEEauthorrefmark{1}\{afentaki, zervakis\}@ceid.upatras.gr,
\IEEEauthorrefmark{4}\{arkokkin, ksiop\}@auth.gr,
\IEEEauthorrefmark{3}\{guerol.saglam, mehdi.tahoori\}@kit.edu
}
}
\maketitle

\begin{abstract}
Printed Electronics (PE) feature distinct and remarkable characteristics that make them a prominent technology for achieving true ubiquitous computing.
This is particularly relevant in application domains that require conformal and ultra-low cost solutions, which have experienced limited penetration of computing until now.
Unlike silicon-based technologies, PE offer unparalleled features such as non-recurring engineering costs, ultra-low manufacturing cost, and on-demand fabrication of conformal, flexible, non-toxic, and stretchable hardware.
However, PE face certain limitations due to their large feature sizes, that impede the realization of complex circuits, such as machine learning classifiers.
In this work, we address these limitations by leveraging the principles of Approximate Computing and Bespoke (fully-customized) design.
We propose an automated framework for designing ultra-low power Multilayer Perceptron (MLP) classifiers which employs, for the first time, a holistic approach to approximate all functions of the MLP's neurons: multiplication, accumulation, and activation.
Through comprehensive evaluation across various MLPs of varying size, our framework demonstrates the ability to enable battery-powered operation of even the most intricate MLP architecture examined, significantly surpassing the current state of the art.
\end{abstract}



\section{Introduction}\label{sec:intro}

Printed electronics (PE) offer a promising solution for introducing computing and intelligence into various domains, including low-end healthcare products like smart bandages, disposables, packaged foods and beverages, smart packaging, in-situ monitoring applications and the vast market of fast-moving consumer goods (FMCG)~\cite{Isohani:Springer:2022,Bleier:ISCA:2020:printedmicro, lacy2020fast}.
These domains impose stringent demands for ultra-low cost (even sub-cent) and conformality, requirements that cannot be met by lithography-based silicon technologies.
On the other hand, PE technology features negligible non-recurrent engineering (NRE) costs, low equipment costs, and ultra-low fabrication cost~\cite{Bleier:ISCA:2020:printedmicro}.
Considering also the inherently supported features of conformality, flexibility, stretchability, non-toxicity, and porosity; PE technology is increasingly recognized as a key enabler for the Internet of Things as part of the ``Fourth Industrial Revolution'', whose core technology advances are functionality and low-cost~\cite{Chang:JETCAS2017}.

By PE we refer to a set of fabrication techniques that are based on printing processes that can realize ultra-low cost, large scale and flexible hardware~\cite{Bleier:ISCA:2020:printedmicro}.
PE does not challenge silicon-based electronics in integration density, area, or speed, mainly due to their large feature sizes arising from low-cost and low-resolution printing.
Typically, operating frequency of printed circuits ranges from a few Hz to only a few kHz~\cite{cadilha2017digital} while their feature size tends to be several microns~\cite{lei2019low}.
On the other hand, due to its form-factor, conformity, low cost, and on-demand, even at low-volume fabrication, PE can target application domains untouchable by silicon VLSI~\cite{Armeniakos:TC2023:codesign}.
However, their large feature sizes and inherent high transistor gate capacitances result in increased power and area compared to nanometer technologies.
Despite the appealing features of PE, such limitations make the realization of complex circuits, as machine learning (ML) classifiers that form the core task in most printed applications~\cite{Armeniakos:TC2023:codesign}, very challenging.

As an attempt to mitigate the aforementioned limitations, the authors in~\cite{Mubarik:MICRO:2020:printedml} exploit the potential for high customization, originating by the low-fabrication and NRE costs of printed circuits and designed bespoke ML classifiers.
The term bespoke refers to fully-customized circuit implementations, tailored to specific ML model and dataset.
The bespoke designs of~\cite{Mubarik:MICRO:2020:printedml} achieved remarkable area and power savings that proved however insufficient towards the realization of complex ML circuits, such as Multilayer Perceptrons (MLPs).
Thus,~\cite{Mubarik:MICRO:2020:printedml} focused only on simple ML models (e.g., Decision Trees).
Targeting more complex printed ML circuits, the authors in~\cite{Armeniakos:DATE2022:axml,Armeniakos:TCAD2023:cross,Armeniakos:TC2023:codesign,Kokkinis:DATE2023} employed Approximate Computing.
Approximate computing for ML circuit design is gaining significant attention since by trading some loss in accuracy, it can achieve high gains in area and power~\cite{AxDNNsurvey, Henkel:ICCAD2022:expedition}.
Though, their proposed approximations are limited in scope and do not exploit the full spectrum of Approximate Computing, resulting in conservative gains.
Similarly,~\cite{Weller:2021:printed_stoch} used Stochastic Computing that resulted however in large accuracy degradation.

In this work, we propose an automated design framework that through means of bespoke design and approximation enables printed-battery powered MLP classifiers.
Unlike the state of the art, we implement a holistic approximation that targets the core components of a MLP circuit: multiplication, accumulation, and activation function. 
Specifically, we use power-of-2 weight quantization to eliminate multiplications and quantized Relu to reduce the size of the outputs of the hidden layer.
Moreover, we propose an accumulation approximation that through a genetic optimization reduces the number of summand bits in each accumulator.
Finally, we also approximate the activation function of the output layer by selectively comparing subsets of its inputs, decreasing, thus, the size of the comparators.
Compared to the state-of-the-art exact baseline, our evaluation shows that, across six MLPs of varying complexity, our framework delivers more than $2.6$x area and $8$x power reduction for less than $5$\% accuracy loss.

\noindent
\textbf{Our novel contributions within this work are as follows}:
\begin{enumerate}[topsep=0pt,leftmargin=*]
\item This is the first comprehensive approximation framework
\footnote{Our framework is available at https://github.com/floAfentaki/Approximation-Techniques-Targeting-Printed-MLPs}
for printed MLPs circuits that apply a holistic approximation across all the MLP components: multiplication, accumulation, and activation function.
\item We propose an activation-aware accumulation approximation, customized for bespoke MLP circuits, that is applied through a multi-objective genetic optimization.
Our proposed area model and accuracy evaluation of approximate printed MLP circuits enable fast and high-level exploration of the corresponding approximation space.
\item Our framework enables printed-battery powered operation of complex printed MLP circuits for up to $5$\% accuracy loss. Specifically, our framework surpasses the current state of the art by increasing the number of parameters that can be integrated into a printed MLP circuit by $20$x.
    
\end{enumerate}
\section{Preamble}
\subsection{Background on Printed Electronics}\label{sec:background}

Moore’s law has been driving the lithography-based silicon VLSI technologies for higher integration density.
However, such technologies are governed by a lower cost bound due to the expensive manufacturing, e.g., wafer processing, lithography, and material processing.
This in turn increases the cost for testing, assembly, and packaging.
PE technology, based on low-cost additive manufacturing technologies, has emerged as an alternative approach that is gaining popularity, especially targeting disposables and domains with ultra-low cost margins, particularly those with conformality requirements. 

Printing technologies commonly utilize mask-less, portable, and additive manufacturing methods.
Such methods can greatly reduce manufacturing costs and decrease production timelines~\cite{chang2017circuits}. 
PE rely on printing processes, such as jet printing, screen or gravure printing~\cite{cui2016printed}.
The simple additive manufacturing and the low equipment costs  enable remarkably low-cost (even sub-cent) electronic circuits.
Though, due to the large feature sizes that result in elevated device latencies and low integration density (orders of magnitude lower compared to silicon VLSI), PE cannot match the area and performance achieved by silicon systems. 
Nevertheless, in the target domains, the performance and precision requirements are typically very low, e.g., sampling rate of only a few Hz and few bits precision~\cite{Henkel:ICCAD2022:expedition}.
Such requirements could effectively be fulfilled by printing technologies under acceptable area and energy constraints.  
In this work we consider the Electrolyte-Gated FET (EGFET) technology that has good supply voltage and mobility characteristics, being thus good fit for battery-powered applications~\cite{Bleier:ISCA:2020:printedmicro}.

\subsection{Related Work}\label{sec:related_works}

 \begin{table}[t]
  \centering
  \renewcommand{\arraystretch}{1.1}
  \caption{Printed MLP Circuits State of the Art}
 \label{tab:comparison_related}

      \begin{tabular}{|c | c|c|c|c| }
      \hline
      \textbf{Works} & \textbf{Bespoke} & \makecell{\textbf{Approx.} \\ \textbf{Multiplication}} &\makecell{\textbf{Approx.} \\ \textbf{Addition}} & \makecell{\textbf{Approx.} \\ \textbf{Activation}}   
  \\ \hline    \hline

\cite{Mubarik:MICRO:2020:printedml} & \greentick & \redx  & \redx  & \redx \\ \hline 
\cite{Armeniakos:DATE2022:axml}, \cite{Armeniakos:TCAD2023:cross} & \greentick & \greentick & \redx & \redx\\ \hline 

       
\cite{Armeniakos:TC2023:codesign} & \greentick & \greentick & \greentick &  \redx      \\ \hline 

\cite{Weller:2021:printed_stoch} & \greentick & \redx  & \redx  & \greentick \\ \hline \hline

     \textbf{Ours} & \greentick & \greentick & \greentick & \greentick   \\ \hline

      \end{tabular}

  \end{table}

In recent years, the design of complex systems based on flexible technologies has gained vast research interest.
Briefly, in 2020 Ozer et al., fabricated a processing system for odour detection on flexible electronics~\cite{Ozer:Nature:2020}.
Weller et al., fabricated a neuromorphic circuit based on the flexible EGFET technology that operates at $1$V~\cite{Weller:ASPDAC:2020}. 
Similarly, in 2021, ARM fabricated the first 32-bit processor on a flexible plastic technology 
~\cite{ARM:2021:PlasticARM}.
In 2023 PragmatIC fabricated ML classifiers with a low area and power footprint on polyamide substrate using the 0.8$\mu$m FlexIC TFT technology~\cite{PragnatIC:TinyMLClassifiers:2023}.
However, these works do not leverage the hardware-efficiency of approximate computing while~\cite{Ozer:Nature:2020, ARM:2021:PlasticARM,PragnatIC:TinyMLClassifiers:2023} do not consider a printed technology.

Design methodologies that aim to shrink the size of neural networks and deploy them on FPGAs at the deep edge are introduced in~\cite{Sung:HPCA2021:Mix&Match,Hanchen:DAC2020:HybridDNN,FixyFPGA2021}.
Although FPGAs support bespoke designs, they feature orders of magnitude higher computing capabilities compared to PE.
Approximating arithmetic blocks for Deep Neural Networks have also been suggested as candidate solutions for the generation of low-power DNNs~\cite{AxDNNsurvey}.
Nevertheless, those methodologies target conventional, non-bespoke implementations and therefore are not suitable for printed applications.

Targeting specifically printed ML classifiers, the authors in~\cite{Armeniakos:TC2023:codesign,Armeniakos:TCAD2023:cross,Balaskas:ISQED2022:axDT,Armeniakos:DATE2022:axml,Mubarik:MICRO:2020:printedml} consider bespoke implementations.
\cite{Mubarik:MICRO:2020:printedml} does not leverage approximate computing while \cite{Mubarik:MICRO:2020:printedml} and~\cite{Balaskas:ISQED2022:axDT} mainly consider simple classifiers.
\cite{Armeniakos:DATE2022:axml} introduced approximate printed ML circuits but approximated only the multiplications and then applied a generic gate-level pruning approximation.
In~\cite{Armeniakos:TCAD2023:cross} the authors extend~\cite{Armeniakos:DATE2022:axml} by applying also voltage over-scaling.
The authors in~\cite{Kokkinis:DATE2023} evaluate the impact of neural compression on printed MLPs but they presented only preliminary results.
Finally,~\cite{Armeniakos:TC2023:codesign} approximates both the multiplication and accumulation.
However,~\cite{Armeniakos:TC2023:codesign} applied only coarse-grain truncation on the accumulators, limiting thus the potential gains.
Table~\ref{tab:comparison_related} summarizes relevant works on printed MLP circuits.
Besides approximate computing techniques, stochastic computing has been suggested as a candidate approach to mitigate the excessive area and power overhead~\cite{Weller:2021:printed_stoch}.
Although the stochastic schemes can yield significant area and power gains, they may also result in a high degradation in the classifier's accuracy~\cite{Weller:2021:printed_stoch} and potentially increased classification latency.

Our work differentiates from the state of the art as it combines the bespoke design paradigm along with a holistic approximation approach that considers approximate multiplication, accumulation, and activation.

\section{Proposed Framework}\label{sec:framework}

\begin{figure}[!t]
\centering
\includegraphics[width=\columnwidth]{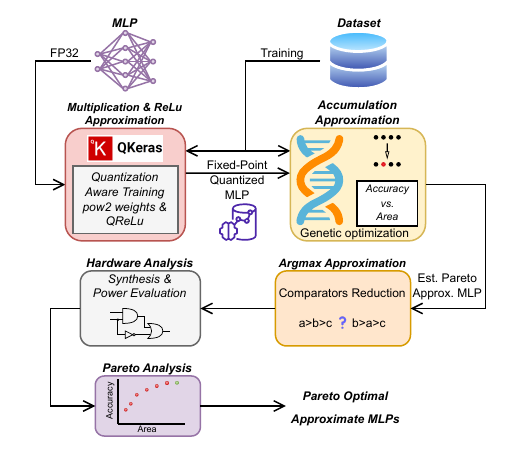}
\vspace{-4ex}
\caption{Overview of our proposed framework.}
\label{fig:framework}
\end{figure}

This section presents our approximation framework (Fig.~\ref{fig:framework}) which aims to minimize the area-overhead of a printed MLP circuit while maintaining high accuracy.
Our framework takes as inputs a trained MLP model and the corresponding train and test datasets. 
Without loss of generality, if the MLP is not pre-trained, it can be trained as described in~\cite{Mubarik:MICRO:2020:printedml,Armeniakos:TC2023:codesign} and Section~\ref{subsec:preliminaries}.
Operating in a fully automated manner, our framework produces a set of area-accuracy Pareto-optimal printed MLP circuits by employing bespoke design and a holistic approximation approach that encompasses the approximation of all components within a MLP neuron, i.e., the multiplication, accumulation, and activation circuits.
Sections~\ref{subsec:axmult} to~\ref{subsec:axadd} describe the approximations applied by our framework while Section~\ref{subsec:flow} describes its overall flow. 

\subsection{Preliminaries}\label{subsec:preliminaries}
The baseline MLPs considered in our work use the same topology as in~\cite{Mubarik:MICRO:2020:printedml,Armeniakos:TC2023:codesign} in order to enable fair comparisons in Section~\ref{sec:experimental}.
The datasets are obtained from the UCI ML repository~\cite{Dua:2019:uci}.
We train the MLPs using scikit-learn and the randomized parameter optimization with $5$-fold cross validation.
Similar to~\cite{Mubarik:MICRO:2020:printedml,Weller:2021:printed_stoch}, the inputs are normalized to $[0,1]$ and we randomly split the datasets to $70$\%/$30$\% train/test.
The Relu activation function is used in the hidden layer.

For the design of the corresponding printed MLP circuit, either approximate or accurate, we employ the efficiency of bespoke design paradigm~\cite{Mubarik:MICRO:2020:printedml}.
For each MLP a fully parallel architecture (i.e., one inference per cycle) is implemented in which the weight values are hardwired in the circuit.
In addition, we follow design optimizations of~\cite{Armeniakos:TC2023:codesign,Kokkinis:DATE2023}.
Since we implement a bespoke design and the input activations of all neurons are positive (e.g., Relu), we split the weights of each neuron to positive and negative ones.
For the negative ones the absolute value is used. 
The respective products are accumulated separately, i.e., two distinct accumulators are used.
Finally, the two obtained sums are subtracted. 
As a result, we almost completely avoid signed arithmetic and the associated hardware overhead of sign-bit extension, etc.

Finally, we truncate the inputs of the MLP down to $4$ bits. 
An input size of $4$ bits is small enough and doesn't result in any accuracy degradation across all the examined datasets.

\begin{figure}[!t]
\centering
\includegraphics[width=\columnwidth]{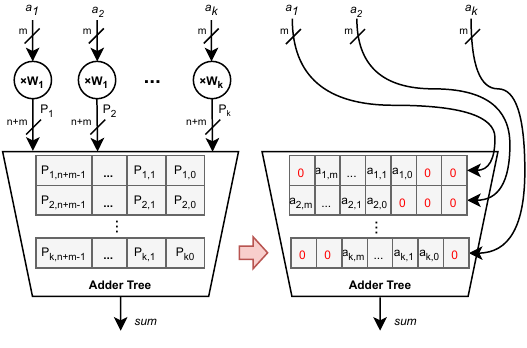}
\caption{Showcase of the impact of power-of-2 weights on a bespoke MAC circuit.
On the left bespoke multipliers and a generic adder tree are used.
With power-of-2 weights, only a simpler and narrower adder tree is required.
}
\label{fig:mac}
\vspace{-2ex}
\end{figure}

\subsection{Multiplication Approximation}\label{subsec:axmult}
The multipliers within a neuron consume the largest part of its area~\cite{AxDNNsurvey}.
Though, in bespoke circuit design (as in our work), the coefficients of a machine learning model, such as the weights of MLPs, are hardwired in the circuit description.
Consequently, the area overhead of a neuron's multipliers is strongly influenced by the values of its weights.
In an effort to leverage this, the state-of-the-art~\cite{Armeniakos:DATE2022:axml,Armeniakos:TC2023:codesign,Armeniakos:TCAD2023:cross} explored custom approaches that replace the MLP weights with more hardware-friendly values (i.e., weights that instantiate smaller bespoke multipliers).
However, even with these modifications, the resulting circuits still require multiplications and the associated hardware overheads remain prohibitive. 

This observation has motivated us to consider that the elimination of the multiplication operation is mandatory towards realizing complex printed MLP classifiers.
To achieve this, we replace the weights with power-of-2 values. 
Since the weights are hardwired in the circuit, power-of-2 weights transform every multiplication to simply wiring.
Thus, the area of multipliers is nullified.
An illustrative example is depicted in Fig.~\ref{fig:mac}.
As shown, in Fig.~\ref{fig:mac} not only the multipliers are removed from the neuron but also a semi-bespoke adder tree is required for the accumulation since several summand bits in the tree (see Fig.~\ref{fig:mac} left) are replaced by constant zero values (see Fig.~\ref{fig:mac} right).
Power-of-2 weight representation has been widely explored to improve ML inference performance but in many cases it is not preferred since it may incur an unacceptable accuracy loss.
However, it is important to acknowledge that in printed applications, the feasibility of a design is the primary concern and takes precedence over achieving the utmost classification accuracy~\cite{Mubarik:MICRO:2020:printedml, Armeniakos:TC2023:codesign}.

We utilize power-of-2 quantization to transform the MLP's weights into powers of two.
Our framework uses Google QKeras~\cite{Coelho:Qkeras}, a tool specifically designed for such purposes, and utilizes its power-of-two (po2) quantizer.
The weight size is set to $8$ bits, which is a commonly used size in neural network quantization~\cite{AxDNNsurvey}. 
Biases are also quantized along with the weights in a similar manner. 
We perform Quantization aware re-training (QAT) using Qkeras
to effectively recover any tentative accuracy degradation due to the po2 quantization. 
Note that the MLP models considered for printed applications are fairly small in size~\cite{Mubarik:MICRO:2020:printedml} (compared to contemporary Deep Neural Networks) and as a result QAT requires only few retraining epochs,
even for the most complex printed MLPs.

\subsection{Activation Approximation}\label{subsec:axact}
To minimize hardware cost, printed MLPs are trained with a single hidden layer.
Similarly, we use the Relu activation function at the hidden layer and Argmax at the output layer.

\subsubsection{QRelu}\label{subsec:qrelu}
While Relu implementation only requires a few AND gates, it should be noted that Relu is unbounded and produces large bit-width outputs as the accumulation of the weights-input activations products is performed in full precision.
Consequently, this results in a significant area overhead at the neurons of the subsequent layer, as they operate over inputs with a large bit-width.
To mitigate this overhead, our framework employs quantized Relu (QRelu) where the quantization size is set again to $8$ bits.
For hardware efficiency, linear QRelu with truncation is used.
The circuit complexity of QRelu remains insignificant since it requires only a few AND gates for nullification and a few OR gates for clipping.
To retain high accuracy, we incorporate the activation quantization  (QRelu) of the hidden layer in QAT.

\subsubsection{Approximate Argmax}
Argmax is the activation function of the output layer and  is implemented as a tree of comparators determining the neuron with the highest value.
Typically, the comparators compare the outputs of the neurons in the order they appear, i.e., \nth{1} neuron with \nth{2} neuron, \nth{3} neuron with \nth{4} neuron, etc.
However, we observe that there are correlations between the neurons' outputs.
For example, we observe that, in most cases, when neuron $e$ has the maximum output, neuron $e$ and neuron $z$ feature so close values that only a few LSBs might be sufficient for an accurate-enough comparison.
Similarly, when neuron $f$ has the maximum output, the difference of neuron $f$ and neuron $g$ is so high that only a few MSBs suffice for a good-enough comparison.

Given this potential for decreasing the size of the required comparators, we approximate the Argmax activation by identifying the appropriate order of comparisons as well as the minimum subset of bits that need to be compared each time.
First, for each neuron $i$ and each neuron $j$ we compare them in an approximate way while the rest comparisons are performed accurately. 
For the approximate comparison we employ a greedy approach to extract the minimum subset of bits that need to be compared so that the classification accuracy (on the train dataset) remains almost the same (i.e., does not drop more than $0.5$\%).
Our greedy approach is straightforward: it starts from the MSB and decides if the corresponding bit should be kept or discarded based on the accuracy obtained without that bit.
After this procedure is over $\forall i, j$, we fill a 2-D matrix that contains the minimum set of bits that will be kept for each comparison.
Finally, we use the Hungarian algorithm~\cite{Hungarian} to select the combination (i.e., which $(i, j)$ will be compared each time) that gives the lowest cost (i.e., smallest number of bits to be compared in total).
Each $i$, $j$ can be selected only once.
The Hungarian algorithm is commonly used in assignment problems such as is our case.
Overall the size of the matrix is fairly small (up to $16\times 16$ for the examined MLPs) so the algorithm advances very fast.
The above procedure is repeated for all subsequent (few) comparison stages.

\subsection{Accumulation Approximation}\label{subsec:axadd}
After QAT,
the weights are in power-of-2, the input activations of each layer exhibit reduced size and semi-bespoke adder trees are required for the accumulation.
Next, to further improve hardware efficiency, our framework approximates the accumulation operation by selectively removing certain summand bits from the adder trees.
Removing a summand bit is equivalent to replacing it by a constant zero in the hardware description of the MLP.
Hence, unlike custom arithmetic approximations that mainly alter the citcuit's logic~\cite{Honglan:JPROC:2020:axsurv}, we fully leverage the IPs and optimization capabilities of the EDA synthesis tool, which among others includes constant propagation, to optimize the obtained circuit even further.

\begin{figure}[!t]
\centering
\includegraphics[width=\columnwidth]{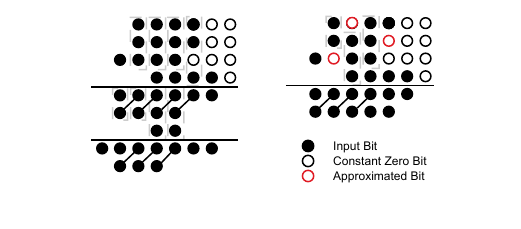}
\vspace{-6ex}
\caption{Example of our implemented accumulation approximation.}
\label{fig:addapprox}
\vspace{-2ex}
\end{figure}

A descriptive example of our accumulation approximation is illustrated in Fig.~\ref{fig:addapprox}.
In this example, the summation of four 4-bit operands is showcased.
The black dots represent input bits, the values of which are not known beforehand, while the white dots indicate zero values resulting from the multiplication by the constant power-of-2 weights.
As shown in Fig.~\ref{fig:addapprox}, the exact addition requires $6$ full-adders, $2$ half-adders, and necessitates three accumulation stages.
In contrast, by selectively removing only three bits (out of $16$), the approximate adder tree reduces the hardware requirements to $2$ full-adders, $1$ half-adder, and eliminates one accumulation stage (i.e., delay gain as well).

The state-of-the-art arithmetic circuit approximation approaches typically focus on approximating a few least significant bits (LSBs) until a certain accuracy threshold is reached~\cite{Honglan:JPROC:2020:axsurv}.
However, this intuitive approach may not always be applicable in our specific case due to the QRelu activation.
As described in Section~\ref{subsec:qrelu}, the hidden layer uses QRelu that truncates certain LSBs of the accumulation result and also applies clipping to a maximum value.
Hence, in that case the middle bits might be more significant than the higher order bits. 
Moreover, due to QRelu (i.e., non-liner function) the impact of removing each bit on the final classification accuracy becomes intricate and challenging to model.
Additionally, the gains of removing a specific bit also depend on its location (column in Fig.~\ref{fig:mac}).
Although removing different bits from the same column may lead to similar hardware gains, this does not necessarily hold true for the classification accuracy due to the different distributions of each input that may affect the probability of a removed bit being zero or one.

As a result, to minimize the hardware overheads of a printed MLP circuit, our framework needs also to identify, for each adder tree in the MLP circuit, which bits shall be removed.
However, our framework must maximize the area gains while also preserving high classification accuracy.
To address this optimization problem we employ a Genetic Algorithm due to its inherent parallelism and its ability to effectively explore the solution space.
Though, other heuristics or optimization techniques may also be employed.

Overall, our accumulation approximation differs from traditional arithmetic approximation~\cite{Honglan:JPROC:2020:axsurv} in several ways.
Firstly, our method is activation-aware as it accounts the configuration of QRelu to identify the more/less important columns for approximation.
Additionally, our approach considers the accuracy of the entire MLP, capturing dependencies or synergies of different approximations.
Furthermore, it is input-aware, as the distribution of each input plays a crucial role in our approach.
For instance, unlike the state-of-the-art arithmetic approximation~\cite{Honglan:JPROC:2020:axsurv}, our approach does not consider different bits in the same column as equivalent for approximation.

\subsubsection{Genetic Optimization}
In our framework, each candidate approximate solution for the accumulation approximation is represented by a set of integers (which we refer to as a "chromosome" from further on) in order to facilitate easy manipulation during the optimization process.
A tentative approximate solution includes all summand bits of each adder tree\footnote{Each neuron uses two adder trees, one for the ``positive'' and one for the ``negative'' products accumulation (see Section~\ref{subsec:preliminaries}).}
in the MLP which can be either removed (value 0) or not-approximated (value 1):
\begin{equation}
\begin{split}
    \mathrm{Candidates} = \{ (b, v): v\in\{0,1\},\,  
     & \forall b \in \mathrm{AddTree}, \\
     & \forall \mathrm{AddTree} \in \mathrm{MLP}\}.
\end{split}
\end{equation}
For example, possible $b$ values are the $a_{i,j}$ in Fig.~\ref{fig:addapprox}, and can be represented by the tuple $(i,j)$ for the specific tree.

To traverse the design space we use the multi-objective Non-dominated Sorting Genetic Algorithm II (NSGA-II)~\cite{Deb:NSGA:2002}. 
NSGA-II receives  the approximation candidates to generate approximate solutions and evaluate them.
Targeting to incentivize the exploration of solutions with high accuracy at the initial stages of evolution, we create an initial population of semi-random chromosomes that are biased towards non-approximated summand bits. 
Our optimization targets two objectives: classification accuracy and area overhead. 
Thus, the obtained approximate solutions will exhibit the most dominant combination of accuracy-area trade-off.
Additionally, we set an upper bound of $15$\% at the accuracy loss  to discourage the exploration of solutions with unacceptably low accuracy. 
Finally, we apply random mutation to the generated chromosomes.

Evaluating the accuracy and area of each candidate solution would typically involve generating its corresponding HDL description and use EDA tools to perform synthesis to get the area and simulation to get the accuracy of the approximate MLP.
However, given the large number of approximate solutions that need to be evaluated in each iteration and the potential licensing constraints of EDA tools, this approach can adversely affect the parallelism and performance of our optimization process or even render it infeasible.
To address this challenge, we employ two high-level methods for evaluating accuracy and estimating area of each approximate solution.

\subsubsection{High-level Accuracy Evaluation}
Obtaining the accuracy of a MLP with our accumulation approximation can be easily implemented at a high-level.
To accomplish this, we have developed a custom MLP classifier class that utilizes the pairs $(b,v)$ from the chromosomes to mask the summands (if a bit is removed the corresponding mask bit is zero).
A bitwise AND between each mask\footnote{The masks are also shifted w.r.t. to the weight values to align the summand and the mask.}
and summand is performed and then addition is just computed on the masked summands.
Weights and inputs are by definition in fixed point representation (quantized inputs and weights) enabling our masking approach.

\subsubsection{High-level Area Estimation}
Evaluating the area, on the other hand, without synthesis is more complex. 
Therefore, we employ a surrogate model to estimate the area overheads of an approximate candidate solution. 
After QAT the multipliers are removed from the circuit and thus, the adders mainly contribute to the overall MLP's area.
Hence, estimating the area of the adder trees can provide a good enough estimation of the overall MLP area.
To achieve this, we assume carry-save operation and for each adder tree in the MLP we count the full-adders (FAs) required for reduction stage.
In other words, for each column in the tree we need to calculate how many FAs are required to reduce the number of summand bits in that column down to two.
Note that a full adder is a 3-to-2 compressor but one of its outputs is of higher order. 
Hence, if $L$ is the number of non-zero bits in a column $\lceil\frac{L-2}{2}\rceil$ FAs are required.
However, we need also to account for the carries coming from the column to the right (each FA gives a carry).
Therefore, for a column $k$, the number of FAs required are:
\begin{equation}
    \mathrm{FA}_k= \lceil\frac{L_k+FA_{k-1}-2}{2}\rceil,\quad \mathrm{FA}_{-1}=0.
\end{equation}
Note that $L_k$ can be easily obtained from the $(b,v)$ values of the corresponding chromosome.
Hence, the total number of FAs is estimated by:
\begin{equation}
\begin{split}
    \mathrm{FA}_{AddTree} &= \sum_{\forall k}\mathrm{FA}_k \\
    \text{and} \quad \mathrm{FA}_{MLP} &= \sum_{\forall AddTree}\mathrm{FA}_{AddTree}.
\end{split}
\end{equation}

\begin{table}[t!]
\setlength\tabcolsep{10pt}
\caption{Spearman's Rank Correlations of Our Area Estimator}
\label{tab:correlation}
\footnotesize
\centering
\renewcommand{\arraystretch}{1.1}

\begin{tabular}{l|c}
\hline
\textbf{Dataset} & \makecell{\textbf{Spearman's Rank}\\ \textbf{Correlation}}  \\ \hline
\textbf{Arrhythmia}      & 0.96 \\
\textbf{Breast Cancer} & 0.96 \\
\textbf{Cardio}      & 0.99 \\
\textbf{Pendigits}   & 0.99\\
\textbf{RedWine}    & 0.96\\
\textbf{WhiteWine}   & 0.98 \\ \hline \hline
\textbf{Average}   & 0.97 \\ \hline

\end{tabular}

\end{table}

Our area model assumes only full-adders and no half-adders.
Moreover, it does not consider a specific reduction strategy (e.g., Wallace, Dadda etc.) that might affect the total number of FAs.
However, for our optimization we do not need an area model that precisely captures the area of an approximate MLP.
We just need a surrogate model that captures accurately enough, the relative order of different accumulation approximated MLPs in order to guide our genetic algorithm towards more area efficient solutions.
In Table~\ref{tab:correlation}, we evaluate the Spearman's rank correlation of our area estimator. 
For each MLP considered (see Section~\ref{sec:experimental}), we run QAT and then we randomly create $1000$ chromosomes and generate the respective approximate MLP circuits (HDL description).
We synthesize the obtained circuits with the EDA tool and measure their area.
Finally, we use our area model to estimate the area of the respective MLP-chromosome combination and calculate the corresponding Spearman Correlation across all designs.
In total, 6000 designs are synthesized for the evaluation of Table~\ref{tab:correlation}.
As shown, our area estimator features almost perfect correlation and thus it is able to efficiently drive our optimization search.
Specifically, it achieves more than 0.96 Spearman correlation while its average value is 0.97.

Overall, our high-level accuracy evaluation and area estimation enable fast exploration of the associated design space.
At the worst case (i.e., Arrhythmia MLP), our genetic optimization requires only $3$h.
The experiments are conducted on an AMD EPYC 7552 with $256$GB RAM.
The population size is set to $1000$ and our genetic run for $30$ iterations.

\subsection{Holistic Approximation Flow}\label{subsec:flow}

In this section, we describe the flow of our framework towards implementing a holistic approximation across all the core components of a MLP.
Overall, as shown in Fig.~\ref{fig:framework}, our framework operates as follows.
First,  applies QAT on the given MLP to approximate the multiplications (power-of-2 weights) and the activations of the hidden layer (QRelu). 
Our accumulation approximation is then applied through the described genetic optimization.
The output of this phase is a set of estimated area-accuracy Pareto-optimal approximated printed MLPs.
Next, for each circuit, our framework approximates the activation function of the output layer (Argmax approximation).
This step is performed last since it leverages and depends on the distribution of the outputs of the output neurons.
The obtained approximation configurations are translated in HDL description and then a hardware analysis is performed on the obtained circuits. 
Finally, a Pareto analysis is performed to extract the designs with the best accuracy-area trade-off. 
In our framework all optimizations are performed on the train dataset while the test dataset is used only for the final assessment of the obtained Pareto-optimal approximate designs.

\section{Results and Evaluation}\label{sec:experimental}
In this section, we present a comprehensive evaluation of our framework. 
First, we analyze the area-efficiency of our implemented approximations.
Then, we compare our framework against the current state-of-the-art printed MLPs~\cite{Mubarik:MICRO:2020:printedml, Armeniakos:TCAD2023:cross, Armeniakos:TC2023:codesign, Weller:2021:printed_stoch}.
Finally, we evaluate the effectiveness of our framework on enabling printed-battery powered MLP classifiers.
We consider the Cardiotocography, Pendigits, Red Wine, White Wine, Arrhythmia, and Breast Cancer datasets as in~\cite{Mubarik:MICRO:2020:printedml, Weller:2021:printed_stoch}.
Synopsys Design Compiler S-2021.06 and VCS T-2022.06 are used for circuit synthesis and simulation respectively, while PrimeTime T-2022.03 is used for circuit simulations. 
All circuits are mapped to the open-source printed EGFET library~\cite{Bleier:ISCA:2020:printedmicro}.
The accuracy numbers reported hereafter regard the test dataset while all designs have been synthesized at a relaxed clock period to improve even further area efficiency.
Specifically, to align with the state of the art, we consider $200$ms for all the datasets except for Pendigits and Arrhythmia that require $250$ and $320$ms, respectively. 
Note that such low clock frequencies are in compliance with typical printed electronics performance~\cite{cadilha2017digital}.
Hereafter, as baseline circuits we refer to the exact bespoke MLP circuits that use $8$-bit fixed point weights and $4$-bit inputs and are designed as in~\cite{Mubarik:MICRO:2020:printedml}.

\subsection{Evaluation of Our Framework}\label{subsec:exp1}
Table~\ref{tab:baselines} presents the topology of each MLP and reports the hardware requirements of the baseline printed MLPs.
As shown, the baseline MLPs feature unbearable area overheads 
($71$cm$^2$ on average) that prohibit realistic application.
Moreover, their power consumption is so high that none of the examined MLPs can be powered by an existing printed power source~\cite{Mubarik:MICRO:2020:printedml}.
In Table~\ref{tab:baselines}, we also present the respective values when we apply QAT only, i.e., eliminate the multipliers with power-of-2 weight quantization and use of QRelu.
As shown, compared to the baseline, when applying QAT the accuracy loss is 1.25\% on average and goes up to $4.4$\%.
On the other hand, for this small accuracy loss, the area gains range from 2.5x up to 5x and the power savings are from 2.5x up to 5.5x. 
Still, despite the impressive gains, the area remains relatively high 
for most MLPs, while only Breast Cancer and Red Wine can be powered by a printed battery (e.g., Molex $30$mW).

\begin{table}[t!]
\setlength\tabcolsep{3pt}
\caption{Evaluation of baseline and power-of-2 quantized printed MLPs}
\label{tab:baselines}
\footnotesize
\centering
\renewcommand{\arraystretch}{1.1}
\begin{threeparttable}
\begin{tabular}{|l|c|ccc|ccc|}
\cline{3-8}
 \multicolumn{2}{c|}{} & \multicolumn{3}{c|}{\textbf{Baseline}} & \multicolumn{3}{c|}{\textbf{QAT Only}}  \\ \hline
  \multicolumn{1}{|c|}{\textbf{MLP}} & \textbf{Topology\tnote{1}} & \textbf{Acc}\tnote{2}    & \begin{tabular}[c]{@{}c@{}}\textbf{Area} \\ ($cm^{2}$)\end{tabular} & \begin{tabular}[c]{@{}c@{}}\textbf{Power}\\ ($mW$)\end{tabular} 
 & \textbf{Acc}\tnote{2}    & \begin{tabular}[c]{@{}c@{}}\textbf{Area} \\ ($cm^{2}$)\end{tabular} & \begin{tabular}[c]{@{}c@{}}\textbf{Power}\\ ($mW$)\end{tabular} \\ \hline
\textbf{Arrhythmia}      & (274,5,16) & 0.620  & 266  & 998   & 0.610 & 92.5   & 258 \\
\textbf{Breast Cancer}  & (10,3,2)   & 0.980  & 12.0 & 40.0  & 0.965 & 4.6    & 16.6 \\
\textbf{Cardio}         & (21,3,3)   & 0.881  & 33.4 & 124   & 0.884 & 8.8    & 34.1 \\
\textbf{Pendigits}      & (16,5,10)  & 0.937  & 67.0 & 213   & 0.893 & 19.5   & 77.3 \\
\textbf{RedWine}        & (11,2,6)   & 0.564  & 17.6 & 73.5  & 0.568 & 3.4    & 13.7 \\
\textbf{WhiteWine}      & (11,4,7)   & 0.537  & 31.2 & 126   & 0.524 & 8.1    & 31.3 \\ \hline
\end{tabular}
\begin{tablenotes}\footnotesize
\item[] 
$^1$ MLP topology. $^2$ Accuracy.
\vspace{-2ex}
\end{tablenotes}
\end{threeparttable}
\end{table}

\begin{figure}[!t]
\centering
\includegraphics{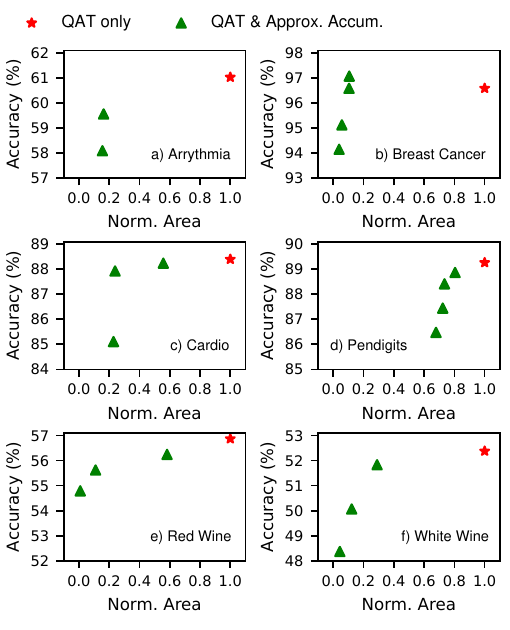}
\vspace{-3ex}
\caption{Evaluation of the effectiveness of our accumulation approximation.
Area is normalized w.r.t. the corresponding QAT-only approximate MLP.}
\label{fig:pareto}
\vspace{-2ex}
\end{figure}

Next, we assess the effectiveness of our accumulation approximation in further reducing the area of printed MLPs.
To do this, we execute our framework without applying the Argmax approximation step.
Fig.~\ref{fig:pareto} illustrates the Pareto-front of the obtained designs (i.e., designs that apply QAT \& accumulation approximation).
The area value is normalized w.r.t. the area of the corresponding \textit{QAT-only} design.
Designs with up to $5$\% accuracy loss w.r.t. the corresponding QAT-only MLP are depicted in Fig.~\ref{fig:pareto}.
As shown, compared to QAT-only, our accumulation approximation achieves 24x area reduction on average for less than $2$\% lower accuracy.
At the worst case (Pendigits at $1$\% lower accuracy), our approximate accumulation reduces the area by $1.3$x.
Fig.~\ref{fig:pareto} demonstrates that applying our accumulation approximation on top of QAT delivers a substantial improvement in area efficiency, without significantly compromising the accuracy of the printed MLPs.

Finally, we examine the additional area reduction that can be achieved when considering also our Argmax approximation.
After eliminating the multiplications and approximating the accumulations, Argmax might occupy a considerable part of the overall printed MLP circuit.
Table~\ref{tab:argmax} presents the impact of applying Argmax approximation on the green points of Fig.~\ref{fig:pareto}, i.e., designs that apply QAT and accumulation approximation.
On each design in Fig.~\ref{fig:pareto} (each green point), we apply our Argmax approximation and compute the area reduction and accuracy loss w.r.t. the initial MLP (green point).
Table~\ref{tab:argmax} presents the average area reduction and accuracy loss for each case.
Moreover, Table~\ref{tab:argmax} evaluates the efficacy of our Argmax approximation in decreasing the size of the required comparators.
Indicatively, if the initial MLP requires $16$-bit comparators while the Argmax-approximated requires $4$-bit comparators, on average, the achieved average comparator size reduction is a $4$x.
As shown, our Argmax approximation reduces the size of the required comparators by $7.6$x on average.
In terms of area, applying our Argmax on the QAT \& approximate accumulation MLPs, reduces the area by an additional $14$\% while the additional accuracy drop is $0.1$\%.

Overall, the above analysis demonstrates that only when applying our holistic approximation we can minimize the area of a printed MLP.
It is noteworthy that applying the state-of-the-art power-of-2 quantization alone is insufficient to enable battery-powered operation of printed MLPs; therefore, our additional approximations (such as the accumulation approximation) are essential in achieving this objective.

\begin{table}[t!]

\caption{Evaluation of Argmax Approximation.}
\label{tab:argmax}
\footnotesize
\centering
\renewcommand{\arraystretch}{1.3}
\begin{threeparttable}
\begin{tabular}{l|c|c |c}
\hline

 \textbf{MLP} & \makecell{\textbf{Avg.} \\ \textbf{Accuracy Loss}\tnote{1}} &  \makecell{\textbf{Avg. Area} \\ \textbf{Reduction}\tnote{1}} & \makecell{\textbf{Avg. Comparator} \\ \textbf{Size Reduction}\tnote{1}} \\ \hline

\textbf{Arrythmia}      &  0.007 & 12\% & 11.4x\\
\textbf{Breast Cancer}  & -0.008  & 21\% & 4.8x\\
\textbf{Cardio}         & -0.001  & 16\% & 6.1x\\
\textbf{Pendigits}      & 0.000  & 9\% & 4.0x\\
\textbf{RedWine}        & 0.007  & 7\% & 11.0x\\
\textbf{WhiteWine}      & 0.002  & 18\% & 8.9x\\ \hline
\end{tabular}
\begin{tablenotes}\footnotesize
\item[] $^1$Values calculated over the respective QAT \& Approximate Accumulation MLP (i.e., green points in Fig.~\ref{fig:pareto}).
\vspace{-2ex}
\end{tablenotes}
\end{threeparttable}
\end{table}




\subsection{Comparison Against the State of the Art}\label{subsec:exp2}
\begin{figure}[!t]
\centering
\includegraphics[width=\columnwidth]{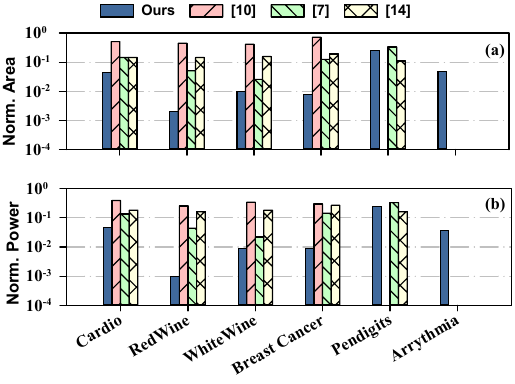}
\vspace{-4ex}
\caption{ (a) Area and (b) power gains of the MLPs generated by our framework compared to state-of-the-art~\cite{Armeniakos:TCAD2023:cross},~\cite{Armeniakos:TC2023:codesign} and~\cite{Weller:2021:printed_stoch}.
All the MLPs feature a $5\%$ accuracy loss from our  baseline~\cite{Mubarik:MICRO:2020:printedml}.
Values are normalized w.r.t.~\cite{Mubarik:MICRO:2020:printedml}.
Y-axis is in logarithmic scale.}
\label{fig:compsoa}
\vspace{-3ex}
\end{figure}

In this section we present a comparative study of our framework against the state-of-the-art works~\cite{Armeniakos:TCAD2023:cross, Armeniakos:TC2023:codesign, Weller:2021:printed_stoch}.
For our framework all approximations are applied, i.e., multiplication, accumulation, and activation approximation.
Fig.~\ref{fig:compsoa} presents the area and power comparison.
All values in Fig.~\ref{fig:compsoa} are normalized over the corresponding value of the respective exact bespoke design~\cite{Mubarik:MICRO:2020:printedml}.
For our circuits and~\cite{Armeniakos:TCAD2023:cross,Armeniakos:TC2023:codesign}, targeting high area efficiency and reasonable accuracy drop, we consider up to $5$\% accuracy loss compared to the baseline~\cite{Mubarik:MICRO:2020:printedml}.
It is important to reiterate that feasibility is the primary requirement for printed ML circuits, prioritizing it over strict accuracy constraints.
Though,~\cite{Weller:2021:printed_stoch} cannot achieve such high accuracy. 
The average accuracy loss of~\cite{Weller:2021:printed_stoch}, for the respective MLPs, is 35\%.
In addition, note that our MLPs,~\cite{Armeniakos:TC2023:codesign},~\cite{Armeniakos:TCAD2023:cross}, and~\cite{Weller:2021:printed_stoch} achieve almost identical performance.
Our MLPs and~\cite{Armeniakos:TC2023:codesign, Armeniakos:TCAD2023:cross} produce one inference result per $200$ms ($250$ms for Pendigits).
The MLPs of~\cite{Weller:2021:printed_stoch} require $220$-$230$ms per inference since they use a stochastic bitstream of length $1024$.

As shown in Fig.~\ref{fig:compsoa}, our framework significantly outperforms~\cite{Armeniakos:TC2023:codesign},~\cite{Armeniakos:TCAD2023:cross} and~\cite{Weller:2021:printed_stoch}.
Specifically, compared to~\cite{Armeniakos:TC2023:codesign}, our MLPs achieve 10x lower area and 12.5x lower power on average.
Similarly, compared to~\cite{Armeniakos:TCAD2023:cross}, our MLPs achieve 96x lower area and 86x lower power on average.
Finally, our MLPs deliver 9x and 11x area and power saving, respectively, compared to~\cite{Weller:2021:printed_stoch}.
As shown in Fig.~\ref{fig:compsoa},~\cite{Armeniakos:TC2023:codesign, Armeniakos:TCAD2023:cross,Weller:2021:printed_stoch} do not consider the Arrhythmia MLP, most probably due to its increased complexity.
As a result, for fairness, the reported average gains exclude Arrhythmia.
Still, our framework achieves very high power and area reduction even for Arrhythmia.
Similarly,~\cite{Armeniakos:TCAD2023:cross} did not consider Pendigits either.
It is noteworthy that our framework demonstrates superior area and power efficiency compared to~\cite{Armeniakos:TC2023:codesign, Armeniakos:TCAD2023:cross} and~\cite{Weller:2021:printed_stoch} across all but one MLPs.
Only for Pendigits the stochastic MLP of~\cite{Weller:2021:printed_stoch} achieves slightly lower power and area than our approximate MLP.
Though,~\cite{Weller:2021:printed_stoch} achieves only $22$\% accuracy while we achieve $89.6$\%.

\subsection{Printed-Battery Operation}
Finally, we evaluate the effectiveness of our framework in generating battery-powered printed MLP classifiers.
Again, we consider the accuracy loss constraint of $5$\% compared to the baseline~\cite{Mubarik:MICRO:2020:printedml} and report in Table~\ref{tab:gains5} the hardware requirements of the Pareto-optimal circuits generated by our framework that satisfy this constraint.
In Sections~\ref{subsec:exp1} and~\ref{subsec:exp2}, for fair comparisons, we considered a voltage supply of $1$V for all our circuits.
However, our approximate MLPs are significantly faster than their exact baseline due to the applied approximations (e.g., multiplication elimination, shorter adder trees, etc.).
As a result, we can decrease the supply voltage of our approximate circuits to achieve even higher power gains. 
Considering that EGFET printed circuits can operate even at $0.6V$~\cite{Marques:Materials:2019} and that printed batteries are customizable in terms of polarity, voltage, shape, etc.,~\cite{PrintedBatteries2018}, we set the voltage supply of our approximate MLPs to the minimum supported value, i.e., $0.6$V, and re-synthesize our designs.
All of our approximate printed MLPs, except for Pendigits, meet the corresponding timing requirement at $0.6$V without any issues.
Due to the smaller delay gain of the approximate Pendigits ($20$\%), re-synthesizing it targeting the $0.6$V library, resulted in a larger circuit (in order to meet the timing requirement) but halved its power consumption also.
As shown, in Table~\ref{tab:gains5} all our approximate MLPs can be powered by a printed battery. 
Arrhythmia and Pendigits can be powered by a Molex $30$mW battery, White Wine and Cardio by a Blue Spark $3$mW battery, while Breast Cancer and Red Wine can be powered by only a printed energy harvester. 
Our MLPs achieve on average 151x lower area and 808x lower power compared to the baseline~\cite{Mubarik:MICRO:2020:printedml}.
Table~\ref{tab:gains5} highlights the effectiveness of our framework.
Our framework enables battery operation of a printed MLP that features $1,450$ parameters (weights).
The largest MLPs that can be powered by the state of the art within a reasonable accuracy loss of $5$\% are the White Wine and Cardio that both feature only $72$ parameters~\cite{Armeniakos:TC2023:codesign}.
Therefore, our framework increased the size of the largest supported MLP by $20$x.

\begin{table}[t!]
\setlength\tabcolsep{3pt}
\caption{Evaluating the Battery Operation of our Printed approximate MLP Circuits for 5\% Accuracy Loss Threshold.}
\label{tab:gains5}
\footnotesize
\centering
\renewcommand{\arraystretch}{1.2}
\begin{threeparttable}
\begin{tabular}{|l|ccc|cc|}
\cline{2-6}
 \multicolumn{1}{c|}{}  & \multicolumn{5}{c|}{\textbf{Our Approximate MLPs}}  \\ \hline

  \multicolumn{1}{|c|}{\textbf{MLP}} &  \textbf{Accuracy}    & \begin{tabular}[c]{@{}c@{}}\textbf{Area} \\ ($cm^{2}$)\end{tabular} & \begin{tabular}[c]{@{}c@{}}\textbf{Power}\\ ($mW$)\end{tabular}  & \makecell{\textbf{Area} \\ \textbf{Reduction}\tnote{1}} & \makecell{\textbf{Power} \\ \textbf{Reduction}\tnote{1}} \\ \hline
 
\textbf{Arrhythmia} & 0.588 & 13.51  & 12.80 &  20x &  78x \\
\textbf{Breast Cancer} & 0.961 & 0.08  & 0.08 &  150x &  500x \\
\textbf{Cardio} & 0.851 & 1.35 & 1.57 &  25x &  79x \\
\textbf{Pendigits} & 0.896 & 25.15 & 26.60 &  2.6x &  8x \\
\textbf{RedWine}  & 0.548 & 0.03 & 0.02 &  587x &   3675x \\
\textbf{WhiteWine} & 0.501 & 0.25 & 0.25 &  125x &  506x \\ \hline
\end{tabular}
\begin{tablenotes}\footnotesize
\item[] 
$^1$ With respect to the corresponding bespoke exact baseline\cite{Mubarik:MICRO:2020:printedml}.
\vspace{-3ex}
\end{tablenotes}
\end{threeparttable}
\end{table}

\section{Conclusion}
With its distinctive characteristics, printed electronics technology emerges as a highly promising solution for introducing computing and intelligence to application domains that have yet to experience significant integration of computing.
This includes the expansive market of fast-moving consumer goods, low-end healthcare products, and disposables, among others.
Though, the large feature sizes in printed electronics hinder the realization of complex circuits.
In this work, we tackle this issue and present an automated framework for generating printed MLP circuits.
Our framework combines the bespoke design paradigm along with a holistic approximation across all the MLP components.
Our evaluation shows that our framework advances the state of the art by enabling printed-battery operation of MLP circuits with $20$x more parameters.

\section*{Acknowledgments}
This work is supported
by the funding programme ``MEDICUS'' of the University of Patras
and by
the European Research Council (ERC).

\bibliographystyle{IEEEtran}
\bibliography{references}

\end{document}